# Using Gray Literature to Influence Software Engineering Curricula

James D Kiper
Computer Science and Software Engineering
Miami University
Oxford OH

Brent Auernheimer
Computer Science
California State University, Fresno
Fresno CA

Simon Sultana
Physical Science
Reedley College
Reedley CA

Gursimran Singh Walia
Computer and Cyber Sciences
Augusta University
Augusta GA

## ABSTRACT

Software engineering (SE) evolves rapidly, with changing technology and industry expectations. The curriculum review bodies (e.g., ACM and IEEE-CS working groups) respond well but can have refresh cycles measured in years. For Computer Science and SE educators to be agile, predictive, and adapt to changing technology trends, judicious use of gray literature (GL) can be helpful. Other fields have found GL useful in bridging academic research and industry needs. GL can be extended to SE to aid faculty preparing students for industry.

We address two questions: first, given the velocity of technical change, do current curricular guidelines accurately reflect industry practice and need for our graduates? Second, how can we track current and emerging trends to capture relevant competencies? We argue a study of the scholarly literature will have a limited impact on our understanding of current and emerging trends and curriculum designers would do well to utilize GL. We close with recommendations for SE educators.

## CCS Concepts

Software and its engineering

Social and professional topics~Professional topics~Computing education~Computing education programs~Software engineering education

## 1 Introduction: Adapting SE Curricula

Software engineering (SE) is driven by changing technology and industry expectations. Conversely, curriculum review bodies (e.g., ACM and IEEE-CS working groups) are typically quite deliberate and take a higher-level view than faculty planning their syllabuses. It should be noted that hysteresis, i.e., not "chattering" between states, is desirable at a different frequency for curriculum as compared to syllabus details. However, curriculum recommendations by our professional societies recognize the importance of industry needs [need reference]:

"A relevant question that industry stakeholders would like answered could be: Our industry requires our employees to have specific knowledge at relevant knowledge levels and several key dispositions."

One of the issues we address is the need to make choices in our curricula to prepare our students for these evolving needs. Textbooks tend to focus on stable foundational material or state-of-the-practice. Scholarly research literature has a longer-term focus. Educators and practitioners need a mechanism for incorporating evolving trends into their courses and SE practices without thrashing about less significant trends in fast-changing fields.

Practitioners in SE and other fields) maintain currency through trade journals, websites, blogs, surveys, technical reports, etc. This collection of sources is called the gray literature (GL).

Gray literature "constitutes material that is published without the academic peer-review processes." [1] It consists of "reports (annual, research, technical, project, etc.), working papers, government documents, white papers, and evaluations." [2] We propose accessing GL to discover important trends in SE and computer science (CS) to inform decisions about what our students know and what their future employers expect them to know. We then use this trend analysis from GL sources to examine current SE curricula to assess whether those curricula adequately prepare our students to understand these trends or to learn and adapt to them in the future.

We recognize the limitations of any forecasts of technology trends. However, we believe our approach and methodology produce a thought-provoking catalyst for the SE education and training community. In this paper, we first review related work and relevant background. As part of this section, we give a more detailed description of the relevant GL. Recently, both the ACM and IEEE Computer Society evolved to use competencies to specify curricular content; these are discussed in section 2.2. We discuss our two research questions in section 3 and describe the methods used in searching the GL in Section 4. We present the results of our investigation and discuss the trends identified in Section 5. Finally, we discuss the application of these resultant trends to SE curricula, including recommendations for modest changes, in Section 6.

## 2 Background and related work

SE innovations make the task of preparing academics and practitioners ever-changing. Certainly, there are fundamental principles and concepts underlying SE that change slowly, but other aspects are more fluid. The challenge for educators is emphasizing principles and evolving knowledge.

### 2.1 Related work

Norton [3] proposed an approach to curriculum development based on input from industry experts and asserted that "tasks, to be performed correctly, demand certain knowledge/skills, tools, and positive worker behaviors." As Balaban and Arnon Sturm pointed out [4], SE as a discipline is typically viewed as being more practically focused. Looking to practitioner experts for input on curriculum is nothing new in our discipline. Goues et al. [5] identified the need to bridge the gap between research and practical advice: "when engineers seek answers to their practical problems, 'perfect' scientific knowledge is not always available. If it's not, engineers accept 'good-enough' evidence: case studies, small-scale experiments, blog posts, or even advice from acknowledged experts."

Hakan Erdogmus advocates for the importance of empiricism while recognizing the value of observation and measurement [6]:

> Empiricism at its heart is supporting decisions through evidence based on data, both observations, and measurements. By observations, I mean occurrences that we can simply record. By measurements, I mean things we can count, calculate, or quantify. Measurements have values, whereas observations have descriptions, possibly including contextual information. Observations provide deeper insight into areas in which measurements serve only as proxies for other constructs. Each kind of data has a place, and empiricism entails collection and use of both kinds.

In this research, we present predictions about trends in technology relevant to SE.

### 2.2 Grey Literature

Likewise, in his CHI'88 keynote, Fred Brooks [7] urged conferences and publications to accept "three nested types of results:" findings, observations, and rules-of-thumb.

Scaffidi and Shaw [11] distinguished between high-ceremony and low-ceremony evidence. They recognized that "many users already rely heavily, though perhaps implicitly, on low-ceremony evidence, and professional programmers also consider such evidence".

Garousi et al. [8] noted that practitioners, including software engineers, lack time, expertise, and access to the research literature. They consult GL they understand and trust, including "trade magazines, online blog posts, survey results or technical reports." Although there is no claim of scientific validity in these sources, there is often very useful summary data about the use of particular tools, techniques, or methods. Zhou [9] identified understanding practitioner views as one of five reasons SE researchers use GL. Trusted practitioners are typical authors of GL who base their conclusions on their professional experience, including application of their academic background

Developers use search engines, websites (such as MDN and StackOverflow), social platforms, and newsletters. As concrete examples of GL, IEEE Software magazine, although publishing peer-reviewed material, sees itself for practitioners [36]:

> IEEE Software's mission is to be the best source of reliable, useful, peer-reviewed information for leading software practitioners—the developers and managers who want to keep up with rapid technology change.

Closer to the other end of the spectrum is Crosstalk, The Journal of Defense of Software Engineering [37]. Interestingly, both venues publish articles written by authors of scholarly articles, repurposed for practice.

We assert that GL is one of the most appropriate sources for predictions of future trends in technology. Jacopo Solani [10] saidbthat GL safely bridges academic research and industry needs. In particular "gray literature is gaining more and more momentum for analyzing and discussing current state-of-practice in software engineering." (p. 12) In a 2021 journal article, after examining the last decade of software engineering research, Kamei et al. found "GL has been essential for bringing practical new perspectives that are scarce in traditional literature." [12] This view was echoed by Ashley in his 2019 Ph.D. dissertation [13] Finally, in the time-honored SE technique of "dogfooding" [18] (or more euphemistically "eating your own cooking"), we actively sought GL materials in writing this paper. <INSERT FIGURE>

### 2,3 Software Engineering Competencies

Parnas made a compelling argument about specifying student outcomes for SE undergraduate programs by focusing on competencies rather than knowledge [14]. He noted efforts to create a body of knowledge in SE are doomed to failure for three reasons; the lack of agreement about which of the large number of concepts are useful, velocity of change of the knowledge base, and knowledge being tied to tools that are available

The ACM and IEEE Computer Society take this competency-based approach in their latest curriculum guide for SE (and other computing fields) [15, 16]. Drawing upon the work of Frezza et al. [17], CC2020 contained this definition:

> Competency composes an expanded perspective on education that augments knowledge (knowing what) with its skilled application (knowing how) motivated by purpose (knowing why) to accomplish a task, an outcome of value.

This definition was built on steering committee member Frezza's definition of competency as "the personal qualities causally related to effective performance in an area of work" that integrate "knowledge, skills, and dispositions in a professional context" [17]. The ACM and IEEE [16] have embraced this definition and used it as a basis to help educators bridge "the product of an education and the needs of professional practice in the workplace."

## 3 Research Questions

Our two research questions are:

RQ#1: What are near-future trends in computing and related technology?
RQ#2: How well do current curricular recommendations for SE prepare graduates for these near-future trends?

## 4. Methods

To address RQ#1, we searched the GL for trends in technology as GL is the first place SE practitioners and managers consult for advice about upcoming training needs and future markets.

In searching the GL for trends in technology, we conducted web searches using search terms such as "What are current trends in software engineering" and "What are emerging trends in technology". We also searched for titles in the "most popular" articles in IEEE Software and Computer magazine. In addition, we queried Google Bard about these computing technology trends. We used Bard rather than ChatGPT since Bard includes more recent data. We did not search for specific trends to avoid biasing the results with our preconceived opinions about trends. As a result, we found 40 relevant and appropriately current sources from which we collected 459 trends (including duplicates).

After duplicates were removed and synonyms and similar trend wording were combined, the result was 102 unique trends. The researchers then ranked these trends by using the number of times each was mentioned as a proxy for its significance. For example, the concept of blockchain was mentioned in twenty-two sources. Our assumption is that a high rank implies greater importance for the future.

To address RQ#2, we analyzed the trends we extracted from the GL with respect to the degree to which SE curricular competencies prepare a graduate for these trends. No one would assert that knowledge of all application areas can be in every SE curriculum. For example, the term blockchain was the fourth-highest ranked future trend in our sampling of the GL. A university or college could include a blockchain course in their curriculum, however it is not a fundamental SE concept but rather an important application area. It is appropriate to ask whether a student studying a curriculum that meets the competency requirements of the CC2020 could learn and apply blockchain concepts and techniques. Would an employer planning to use blockchain technology be comfortable hiring a SE graduate with those competencies?

Our analysis is based on our professional judgments. The authors have over 100 years of combined experience in higher education SE and experience in industry and entrepreneurship.

## 5 Results

We addressed RQ#1 by sampling the GL. The results of this analysis are presented in Table 1. The first column gives the trend, and the second column gives the frequency this trend appeared in the 102 sources; the third column lists the percentage of the sources in which this concept appeared.

We restrict our analysis to the trends that appear in at least 10% of the GL sources we sampled. Others with lower percentage include digital twins, progressive web applications, big data, and DevSecOps.

Table 1: Trends in Technology

| Trend | Frequency | Percentage |
| --- | --- | --- |
| AI & ML | 48 | 47.1% |
| cybersecurity | 34 | 33.3% |
| IoT | 28 | 27.5% |
| blockchain | 22 | 21.6% |
| quantum computing | 20 | 19.6% |
| AR & VR | 20 | 19.6% |
| cloud computing | 18 | 17.6% |
| low-code | 16 | 15.7% |
| robotics | 12 | 11.8% |
| edge computing | 11 | 10.8% |

For RQ#2, we analyzed these ten trends in technology with respect to the competencies listed for SE in the ACM CC2020 guidelines [16]. The 15 categories of competencies are software requirements, design, construction, testing, sustainment, process and life cycle, systems engineering, quality, security, safety, configuration management, measurement, human-computer interaction, project management, and behavioral attributes.

## 6 Discussion

Several topics we identified (machine learning (ML) and artificial intelligence (AI) emerging technologies, robotics, cloud computing, and blockchain) were common to those identified by Singh et al. [19] in a survey of workshop participants about courses they intended to study or teach.

The ACM and IEEE curriculum recommendations [16] mentioned several of these topics and called out cybersecurity and data science as current and emerging technologies covered by previous curriculum guidelines. However, they mention that while IoT, cloud computing, AI, and high-performance computing have become increasingly important, they have not yet earned recognition as standalone academic disciplines. Finally, in their discussion on the emerging areas of computing, they also identified digital experience (e.g., 5G), ambient computing (i.e., AR/VR), cognitive technologies (i.e., AI), blockchain, robotics, quantum computing, and data privacy.

In the following subsections, we analyze the ways that the first seven of the trends that we have identified from the GL are or are not supported by the standard curricular recommendations.

## 6.1 Artificial Intelligence and Machine Learning

We combined AI and machine learning (ML) as the latter is a subset of the former [20]. This topic was identified in 47.1% of the sources reviewed, which was the most common reference. Several niversities offer programs dedicated to teaching AI [21], and others offer certificate programs [22]. The recent success of generative large language AI systems such as ChatGPT and Google Bard created a tsunami of interest among a wide variety of industries.

Although not mentioned explicitly in the SE competencies, several support the ability to learn and use AI and ML. Software design competency 2.2 requires evaluating and comparing tradeoffs from alternative design possibilities. Software construction competency 3.2 relates to evaluating a software system against modern software practices. ML is one of those. The increasing use of AI and ML in many fields means the competency to work with engineers and developers from other disciplines (software systems engineering competency 7.4) is relevant here.

## 6.2 Cybersecurity

Cybersecurity or security was cited in a third of the sources analyzed. While there is no doubt the importance of security as a component of SE skills continues to increase, questions arise as to the extent and manner of their coverage. The ACM and IEEE emphasized cybersecurity in Computer Science Curricula 2013 [25] and were involved in the Joint Task Force on Cybersecurity Education in developing Cybersecurity Curricula 2017 [26]. The latter's intent included the goals of developing "comprehensive and flexible curricular guidance in cybersecurity education and (producing) a curricular volume that structures the cybersecurity discipline and provides guidance to institutions".

Blair et al. [27] proposed an approach to integrating security in curricula via a five-step process, including identifying principles, mapping these in the current curriculum, identifying gaps, addressing gaps, and connecting and articulating throughout the curriculum [27]. "It is likewise important to assess these skills as demonstrated by Buckley, Zalewski, and Clarke [28] and their use of pre- and post-test evaluation..

The SE competencies include seven specific software security competencies (competencies 9.1 - 9.7), ranging from applying a security lifecycle model to using secure coding standards to using test cases specific to security. There are four additional competencies (10.1-10.4) related to software safety. The CC2020 SE competencies include an emphasis on security.

## 6.3 Internet of Things

The Internet of Things (IoT) is significant in software development. Peter Newman opined that "the continued growth of the IoT industry will be a transformative force across all organizations." [29]. He predicted the IoT market will grow to over $2.4 trillion, and the number of IoT devices from eight billion to 41 billion, by 2027.

One CC2020 competency requires the ability to describe "system engineering concepts and activities to identify problems or opportunities, explore alternatives, create models, and test them." Another asks SE graduates to "develop the big picture of a system in its context and environment to simplify and improve system architectures for supporting system designers." Although there is not an explicit mention, the inclusion of IoT can be inferred from its ubiquity.

## 6.4 Blockchain

The term blockchain does not appear in the competency list, nor do we assert it should. The question we address for this concept and the others is whether the SE competencies provide the background and depth needed for a graduate to become a productive team member in a team developing or using blockchain technology. The software design competency category includes the ability to "produce a high-level design of specific subsystems that is presentable to a non-computing audience by considering architectural and design patterns." We infer that students would learn about architectural and design patterns and would be familiar not only with specific patterns but also with the pattern concept, value, and use. Blockchain architecture is currently atypical in standard lists of such architecture and design patterns. With the increased importance of blockchain, this situation could change.

Understanding blockchain technology requires knowledge and appreciation for randomness and statistical probabilities. There are no specific competencies for these in the ACM and IEEE curriculum recommendations. However, we can infer some mathematical and statistical sophistication that is typical of most SE majors. In particular, competencies like those in the Software Quality category imply this level of related knowledge: 8.4 "Explain the statistical nature of quality evaluation when performed on software execution."

## 6.5 Quantum Computing

Quantum computing (QC) introduces a unique paradigm whose impact is the subject of research and development. Current QC coding processes [38] have differ from common programming practices. These are applications of linear algebra in the manipulation of quantum gates that are represented by unitary matrices of complex numbers. Introduction of courses about QC into CS and SW curricula will require significant retraining of faculty members. However, the potential computational power of QC, and industry trends. make it imperative that faculty and students are prepared.

## 6.6 Augmented/Virtual Reality (AR/VR)

The consumer market for AR or VR devices is accelerating. Data from the consulting firm Linchpin indicates that, by the end of 2018, 8.9 million units had been sold. [30] By the end of 2022, this number is expected to grow to 65.9 million, with a market value estimated at $3.5 billion.

The analysis of AR/VR technologies is analogous to that of IoT. These are technologies of growing importance whose awareness is increasingly important for software engineers.

## 6.5 Cloud Computing

The implementation of cloud-based technologies was documented in 17.6% of the sources. The CC2020 guidelines include a competency focused on the software construction of a medium-sized cloud-based system. The Software Construction competencies (see Appendix) include "3.3 Develop a distributed cloud-based system". Shanahan and Marghitu [31] described introducing cloud computing concepts as early as an introductory K-12 camp course. The ACM and IEEE [32] included the topic in a knowledge area in their previous curriculum recommendations for SE. A lack of emphasis on cloud computing in North Carolina community colleges was noted by Philips et al. [33].

The availability of the Google cloud, AWS, Microsoft's Azure, and similar publicly accessible cloud platforms enable cloud computing courses.

## 7 Conclusion

In addressing our first research question (RQ#1), we suggested that GL is valuable for knowledgeable predictions of future trends in technology. We then used the SE competencies of CC2020 to critique the support they provided for future trends. As a result of this analysis, we have curricular recommendations:

- Graduating students will undoubtedly need some background in AI and ML in the future to understand how these tools can be leveraged in the workplace.
- Include blockchain architectures in SE courses about patterns.
- An introduction to ML analysis and database access methods should be a part of SE curricula.
- SE curricula should include at least an introductory course about cybersecurity, or else appropriate cybersecurity topics should be explicitly included across the curriculum.
- In support of the existing SE competency 3.3, SE educators should ensure students have experience in distributed cloud-based systems.

As a disclaimer, consider the following observations about software planning and prediction from Raccoon and Dog [34]:

> We believe that most software plans are primarily rationalizations for what the planners wanted to do anyway, rather than well-considered evaluations or comparisons of alternate solutions. For example, many designs are elaborate euphemisms for "I want to use technology x," where x is python, mobile, cloud, analytics, or the buzzword du jour... Rationales must never be naively accepted as meaningful.

Have we SE educators done something analogous with SE curricula? Our research here attempts to avoid this pitfall by collecting data about future trends to help identify gaps in SE curricula. We recognize the limitations of any forecasts of technology trends. However, we believe our approach and methodology have can be a thought-provoking catalyst for the SE education and training community. More importantly, the process described is reproducible and can be done faster than a systematic literature review of peer-reviewed publications.

## REFERENCES

1 V. Garousi and A. Rainer, "Gray Literature Versus Academic Literature in Software Engineering: A Call for Epistemological Analysis," in IEEE Software, vol. 38, no. 5, pp. 65-72, Sept.-Oct. 2021, DOI: 10.1109/MS.2020.3022931.
2 Grey Literature, Wikipedia url: https://en.wikipedia.org/wiki/Grey_literature
3 Norton, R. E. Quality instruction for the high-performance workplace: DACUM. 1998. Retrieved from http://files.eric.ed.gov/fulltext/ED419155.pdf
4 Mira Balaban and Arnon Sturm. 2018. Software engineering lab: an essential component of a software engineering curriculum. Proceedings of the 40th International Conference on Software Engineering: Software Engineering Education and Training (ICSE-SEET '18). Association for Computing Machinery, New York, NY, USA, 21–30. DOI:https://doi.org/10.1145/3183377.3183395
5 C. L. Goues, C. Jaspan, I. Ozkaya, M. Shaw, and K. T. Stolee, "Bridging the Gap: From Research to Practical Advice," in IEEE Software, vol. 35, no. 5, pp. 50-57, September/October 2018, DOI: 10.1109/MS.2018.3571235.
6 H. Erdogmus, "How Important Is Evidence, Really?," in IEEE Software, vol. 27, no. 3, pp. 2-5, May-June 2010, DOI: 10.1109/MS.2010.75.
7 Brooks F. P. 1988. Grasping reality through illusion—interactive graphics serving science. In Proceedings of the SIGCHI Conference on Human Factors in Computing Systems (CHI '88). Association for Computing Machinery, New York, NY, USA, 1–11. DOI:https://doi.org/10.1145/57167.57168
8 Garousi V., Felderer M., Mäntylä M.V., Rainer A. (2020) Benefitting from the Grey Literature in Software Engineering Research. In: Felderer M., Travassos G. (eds) Contemporary Empirical Methods in Software Engineering. Springer, Cham. https://doi.org/10.1007/978-3-030-32489-6_14
9 Zhou, X. (2020). How to treat the use of grey literature in software engineering. Proceedings of the International Conference on Software and System Processes. ACM Digital Library. https://dl.acm.org/doi/10.1145/3379177.3390305
10 Soldani, J. (2019). Grey literature: A safe bridge between academy and industry? ACM SIGSOFT Software Engineering Notes, 44(3), 11-12. https://dl.acm.org/doi/10.1145/3356773.3356776
11 C. Scaffidi and M. Shaw, "Toward a Calculus of Confidence", Proc of 1st International Workshop on Economics of Software and Computation, 2007, pp.7-9.
12 Kamei, F., I. Wiese, Lima, C., Polato, I., Nepo,uceno, V., Ferreira, W., Ribeiro, M., Pena, C., Pinto, G., Soares, S. Grey Literature in Software Engineering: A critical review. Information and Software Technology Volume 138, October 2021. https://www.sciencedirect.com/science/article/abs/pii/S0950584921000860
13 Williams, Ashley. Finding high-quality grey literature for use as evidence in software engineering research. (2019). Ph.D. Dissertation. The University of Canterbury. https://ir.canterbury.ac.nz/handle/10092/17908
14 D. L. Parnas, "Software Engineering: A Profession in Waiting," in Computer, vol. 54, no. 5, pp. 62-64, May 2021, DOI: 10.1109/MC.2021.3057685.
15 Impagliazzo, J., Bourque, P., Mead, N. Incorporating CC2020 and SWECOM Competencies into Software Engineering Curricula: A Tutorial, 32nd International Conference on Software Engineering Education and Training. pp. 12-14, 2020.
16 Computing Curricula 2020 CC2020 Paradigms for Global Computing Education, December 31, 2020, https://www.acm.org/binaries/content/assets/education/curricula-recommendations/cc2020.pdf
17 Frezza, S, Daniels, M., Pears, A., Cajander, A., Kann, V., Kapoor, A., McDermott, R., Peters, A-K., Sabin, M., and Wallace, C. Modelling Competencies for Computing Education beyond 2020: A Research Based Approach to Defining Competencies in the Computing Disciplines. In Proceedings Companion of the 23rd Annual ACM Conference on Innovation and Technology in Computer Science Education, 27. (ACM, 2019).
18 Harrison, W. Eating your own dog food. IEEE Software. May/June 2006, pp. 5-7, vol. 23. https://doi.org/10.1109/MS.2006.72
19 Singh, P., Farooq, S. U., Tiwari, S., and Sureka, A. (2018). An experience report on the workshop on emerging software engineering education. ACM SIGSOFT Software Engineering Notes, 43(2), 12-23. https://doi.org/10.1145/3203094.3203112
20 Burnham, K. (2020, May 6). Artificial intelligence vs. machine learning: What's the difference? Northeastern University Graduate Programs. https://www.northeastern.edu/graduate/blog/artificial-intelligence-vs-machine-learning-whats-the-difference/
21 What is a Bachelor of Software Engineering in Artificial Intelligence? Torrens University. https://www.torrens.edu.au/courses/design/bachelor-of-software-engineering-artificial-intelligence
22 Stand Apart. Machine Learning Certificate from Cornell University.

https://online.cornell.edu/machine-learning

23 Shein, E. How AI is changing software development. ACM News. January 26, 2017. https://cacm.acm.org/news/212583-how-ai-is-changing-software-development/fulltext

24 Dsouza, M. . 5 ways artificial intelligence is upgrading software engineering. PackT Hub. September 2, 2018. https://hub.packtpub.com/5-ways-artificial-intelligence-is-upgrading-software-engineering/

25 Computer Science Curricula 2013 Curriculum Guidelines for Undergraduate Degree Programs in Computer Science, Dec 20, 2013, https://www.acm.org/binaries/content/assets/education/cs2013_web_final.pdf

26 Joint Task Force on Cybersecurity Education. (2017). Cybersecurity Curricula 2017. Technical Report. ACM, IEEE-CS, AIS SIGSEC, and IFIP WG 11.8. https://doi.org/10.1145/3184594

27 Jean R. S. Blair, Christa M. Chewar, Rajendra K. Raj, and Edward Sobiesk. 2020. Infusing Principles and Practices for Secure Computing Throughout an Undergraduate Computer Science Curriculum. Proceedings of the 2020 ACM Conference on Innovation and Technology in Computer Science Education (ITiCSE '20), June 15–19, 2020, Trondheim, Norway. ACM, New York, NY, USA

28 Buckley, I. A., Zalewski, J., and Clarke, P. J. (2018). Introducing a cybersecurity mindset into software engineering undergraduate courses. International Journal of Advanced Computer Science and Applications, 9(6), 448-452. https://pdfs.semanticscholar.org/fa9b/ae6f2307dcf738dbfbc6c0a4f53fbeffbb1c.pdf

29 Newman, Peter, The Internet of Things 2020: Here's what over 400 IoT decision-makers say about the future of enterprise connectivity and how IoT companies can use it to grow revenue. Mar. 6, 2020. https://www.businessinsider.com/internet-of-things-report

30 Trends Transforming the Augmented And VR Industry In 2021, March 3, 2021. https://linchpinseo.com/trends-in-augmented-virtual-reality/

31 Shanahan, J. and Marghitu, D. (2013). Software engineering Java curriculum with Alice and cloud computing. In Proceedings of Alice Symposium on Alice Symposium, 4, 1-6. https://doi.org/10.1145/2532333.2532337

32 Software Engineering 2014 Curriculum Guidelines for Undergraduate Degree Programs in Software Engineering, https://www.acm.org/binaries/content/assets/education/se2014.pdf

33 Philips, J., Fenwick, N., Davis, S., & Tabrizi, N. (2020). Improving student success through an articulation program in software engineering. From Proceedings of the 21st Annual Conference on Information Technology Education, 378-383. https://doi.org/10.1145/3368308.3415358

35 Raccoon, Dog. 2014. The best laid plans of mice and men. SIGSOFT Softw. Eng. Notes 39, 2 (March 2014), 7–14. DOI:https://doi.org/10.1145/2579281.2579286

36 Developer's research process https://www.silvestar.codes/articles/developer-s-research-process/

37 IEEE Software: Building the community of leading software practitioners https://www.computer.org/csdl/magazine/so/about/14208?title=About&periodical=IEEE%20Software

38 Crosstalk, The Journal of Defense Software Engineering https://community.apan.org/wg/crosstalk/

39 Qiskit https://qiskit.org